# Enhancing Productivity: The Role of Management Practices

Peer-Olaf Siebers[1], Uwe Aickelin[1], Giuliana Battisti[2], Helen Celia[3], Chris Clegg[3], Xiaolan Fu[4], Rafael De Hoyos[5], Alfonsina Iona[2], Alina Petrescu[4] and Adriano Peixoto[6]

[1] School of Computer Science & IT, The University of Nottingham
[2] Business School, University of Aston
[3] Business School, University of Leeds
[4] Department of International Development, Queen Elizabeth House, University of Oxford
[5] World Bank
[6] Institute of Work Psychology, University of Sheffield

For correspondence, please contact Uwe Aickelin, uwe.aickelin@nottingham.ac.uk, +44 115 9514215, School of Computer Science & IT, The University of Nottingham, Jubilee Campus, Wollaton Road, Nottingham, NG8 1BB, UK.

# Enhancing Productivity: The Role of Management Practices

**Abstract**

There is no doubt that management practices are linked to the productivity and performance of a company. However, research findings are mixed. This paper provides a multi-disciplinary review of the current evidence of such a relationship and offers suggestions for further exploration. We provide an extensive review of the literature in terms of research findings from studies that have been trying to measure and understand the impact that individual management practices and clusters of management practices have on productivity at different levels of analysis. We focus our review on Operations Management (OM) and Human Resource Management (HRM) practices as well as joint applications of these practices. In conclusion, we can say that taken as a whole, the research findings are equivocal. Some studies have found a positive relationship between the adoption of management practices and productivity, some negative and some no association whatsoever. We believe that the lack of universal consensus on the effect of the adoption of complementary management practices might be driven either by measurement issues or by the level of analysis. Consequently, there is a need for further research. In particular, for a multi-level approach from the lowest possible level of aggregation up to the firm-level of analysis in order to assess the impact of management practices upon the productivity of firms.

**Introduction**

The persistent productivity gap between European countries and the USA has been a recurrent topic in extant research. Moreover, a wide-ranging plethora of indices is used to identify and measure this gap. At first glance and without careful scrutiny, many findings are ambiguous and even contradictory. A great deal of these apparent discrepancies are accounted for by the different metrics and time spans used, the sector that has been focused on and methodological differences in national account procedures. On the other hand, there are a number of reasons why measured productivity may differ, which do not necessarily reflect underlying differences in productivity (Griffith and Harmgart, 2005). Accepting that there is indeed a difference in productivity between different entities, the question remains what causes this difference? One potential answer is the use of different management practices. In this paper, we investigate this issue and review the potential role management practices have on productivity.

## *Management Practices*

Studies that investigate the link between management practices and productivity have assessed the impact of an individual practice in isolation, the effects of joint adoption of practices and the impact of clusters or systems of complementary practices. In this review, we investigate OM and HRM management practices. OM practices focus on systems management and include Information and Communication Technology (ICT), Just In Time (JIT), Total Quality Management (TQM), and lean production, amongst others. HRM practices focus on people management, in particular the recruitment, development and management of employees (Wood and Wall, 2002). Typical HRM practices involve training, development, empowerment and teamwork.

Wall and Wood (2005) suggest it is unlikely that there exists a 'one size fits all' set of productivity-enhancing management principles or practices. Edwards *et al* (2004) builds upon this contingency approach, stating that the success of management practices are firm-specific and these are affected by the prevailing institutional environment.

This section presents an overview of recent studies found in a systematic literature that investigate the link between management practices and productivity/performance. Using the EBSCO database specific keywords were searched for including only journal articles of the last 10 years. These keywords comprised human resource management practices, operations management practices, supply chain partnering, total quality management, team working, business process engineering, empowerment, payment and reward system, performance appraisal and review, employment development, lean thinking, training, target systems and lean production. A summary of the number of papers found per keyword or in some cases keyword combination can be found in Table 1. The search generated 548 hits and the core papers central to this study are reviewed here in detail. The majority of articles revealed by the database searches involved empirical research in the manufacturing sector in particular. The next most popular sector, although notably less prevalent, being the service sector.

With respect to the level of analysis, the vast majority of papers investigate the link between management practices and productivity at firm-level or industry-level, whilst fewer papers have focused their analysis at either plant-level or establishment-level. It is reasonable to think that results from these studies may be arguable to some extent. This is due to the wide range of subjective definitions and measurements of both management practices and performance or productivity, which makes comparisons difficult.

| Keywords or phrases (used in combination with Productivity) | Articles per keyword |
|---|---|
| Management Practices , Productivity Gap | 71-80 |
| Just-In-Time, Outsourcing | 61-70 |
| ICT, Total Quality Management | 51-60 |
| Empowerment | 41-50 |
| Performance and Management Practices | 31-40 |
| Technology and Retail, Retail Productivity, Lean Production | 21-30 |
| BPR, Technology and Management Practices | 11-20 |
| Operational Practices, Supply Chain Partnering, Payment and Reward System, Employment Development, Target Systems, HRM Practices, Performance Appraisal and Review, Lean Thinking, Training and Retail, Team Working, Performance and Retail, Training and Mana | 1-10 |
| UK Productivity Gap, Leaning Culture, Stock Option Scheme, Incentive Pay Scheme, High Performance System Work, Flexible Staffing Levels, Kan-Ban Systems | 0 |
| **Total** | **548** |

Table 1: Number of papers found in the EBSCO database

*Measuring Management Practices*

There is no consensus in the literature on how to measure management practices. The only commonality shared by all the studies is that management practices are measured in a multidimensional fashion. Because of the inherently intangible nature of management practices, it is very challenging to apply objective forms of measurement. Measures are aggregated to facilitate analysis at the plant-, firm-, industry- or country-level.

In the academic literature, these practices are measured using any combination of a variety of scientific methods: self-reported questionnaires, interviews and observations. Questionnaires and interviews may collect data regarding retrospective or concurrent (or less frequently the prediction of future) management practices. The majority of studies conducting empirical research obtain information by surveying a single knowledgeable individual from each unit of interest, and a minority involve more than one respondent. Less frequently, research studies rely on various unstructured assessment methods, such as observations and analysis of field data collected (Rotab Khan, 2000) and observations alone (Arbós, 2002).

Indeed, the most popular and cost-effective method of collecting empirical data from a large sample is, to remotely (usually postal) conduct a questionnaire survey. Another common method is to derive assessments of management practices from structured or semi-structured interviews, whether by telephone or in person. Respondents may be any combination of senior management, Human Resource (HR) managers, workplace representatives or the employees themselves.

Sometimes the method of data collection needs to be tailored to cultural requirements. A study assessing management practices in identified Japanese subsidiaries in both the USA and Russia made a special effort to set actual interview times with organisations in Russia only to talk respondents through the questionnaire (Park *et al*, 2003). This was necessary because Russian organisations are traditionally very protective of company information, and therefore require direct assurances to be willing to share this with people external to the organisation.

There are already established questionnaire measures of particular management practices, and most studies chose to utilise these in their original or a modified format. Typically only in the absence of a suitable existing tool do researchers choose to develop their own instrument.

*ICT and Productivity*

Recent economics and management science research is increasingly focusing on the impact of ICT on productivity. ICT usage permeates virtually every sector of modern economies, and for decades, the world IT sector has been experiencing significant growth with especially enhanced levels of diffusion during recent years. This revolution is rooted in the swift development of ICT as well as in declining prices for its use.

The most common default hypothesis in ICT studies is to expect a positive correlation between the adoption or wider diffusion of ICT and productivity. Yet, the empirical evidence is mixed, with firm-level studies reporting a positive or no productivity effect, while some industry-level studies even find a negative impact.

At firm level, discrepant results can be illustrated by contrasting the findings from Swamidass and Winch (2002)'s comparative study between USA and UK ICT investment, with those obtained by Licht and Moch (1999) in Germany. Both studies include in their analysis manufacturing plants, although Licht and Moch (1999)'s large establishment sample also includes establishments in the service sector. Swamidass and Winch (2002) use descriptive statistical analysis to compare ICT investments and show that the extent of ICT usage has a positive impact on productivity, with higher levels of computerisation in the USA than in the UK being translated in higher productivity and return on investment in the USA than in the UK. However, Licht and Moch (1999)'s analysis based on a Cobb-Douglas production function and OLS estimators fails to find a relation between ICT investment and increases in labour productivity.

Furthermore, at industry-level some studies show that ICT may even have a negative impact on productivity. As such, Wolf (1999)'s study of the service sector finds that higher levels of computerisation - i.e. the office, computing and accounting equipment made available to employees – have lead to lower TFP. This somewhat unconventional result might be explained by the high reliance of the service sector on the quality of the labour input and quality being hard to measure, whereas it is relatively easier to measure the quantitative work improvements brought in by computerisation. In contrast to Wolf (1999), Basu *et al* (2003) suggest that lower levels of ICT investment played an important role in the resulting slowdown of UK productivity growth during the latter half of the 1990s. Their OLS regression results show a positive impact on TFP which is used as a measure of productivity alongside labour productivity. Moreover, O'Mahony and Van Ark (2005) conduct a comparative study on the UK, the USA, Germany and France from 1995 to 2000 and find that the productivity of the UK retail trade sector was responding positively to ICT adoption and diffusion.

The papers by Stiroh (2002), O'Mahony and Robinson (2003), and Vijselaar and Albers (2004) use data at industry-level to estimate the relationship between labour productivity/TFP and ICT. Stiroh (2002) uses the DID estimator to account for the productivity differentials between ICT-using firms and non ICT-using firms. O'Mahony and Robinson (2003) take a more conventional approach including ICT as an extra factor of production in the TFP calculations.

Vijselaar and Albers (2004) estimate the relationship between the increase in ICT using and producing sectors and aggregate TFP performance. The main result coming from these three papers is that, although ICT has a positive correlation with TFP, there is not enough evidence supporting the view that the increase in ICT investment is the reason behind the rise in USA productivity during the second half of the 1990s. Survey papers by Visco (2000) and Pilat (2004) support these findings.

Basu *et al* (2003) challenges the view that ICT has no spillover effects and therefore cannot contribute to explaining differences between US and UK productivity levels. They argue that investment in ICT has a lagged effect upon TFP and that contemporaneous investment in ICT can even have a negative effect upon TFP. Taking data for the whole US economy at industry-level, they found that growth in ICT between 1980 and 1990 has had a positive effect upon TFP growth between the years 1995 and 2000. Conversely, growth in ICT between 1995 and 2000 has been negatively correlated with growth in TFP during the same period. For the UK, the evidence was not conclusive: lagged ICT growth has not affected the present TFP growth, although present ICT growth was negatively related with TFP growth. Given that the UK investment in ICT during the 1980s was lower than the ICT investment in the USA, the lagged effect of ICT growth upon TFP growth could at least partly explain the relatively lower productivity levels in the UK.

The corroboration of the mixed findings from the literature surveyed above - in particular with regard to finding little or negative productivity impact of what is often an expensive and time-consuming fundamental organisational change - needs not deflate enthusiastic public and private initiatives aimed at encouraging ICT adoption and diffusion. For the answer to harnessing the potential for productivity growth lies in complementary or joint practices that mediate the effects of ICT. Recent studies[1] have highlighted the potential synergistic effects obtained by combining ICT with complementary management practices such as firm reorganisation, innovations in production organisation, product design or the recruitment of skilled labour. For instance, Black and Lynch (2001) analyse labour productivity in panel and cross-section data from 1987 to 1993 on 600 manufacturing USA firms. Using a Cobb Douglas production with Within Group and GMM estimators, they find that ICT diffusion (measured as computer usage by non-managerial employees) combined with workplace reorganisation leads to higher labour productivity. However, this productivity increase is mediated by how workplace reorganisation is implemented, and especially by the level of education and worker training. Skill levels and IT are also found to be complementary (alongside new work organisation and new products and services) in the study by Bresnahan *et al* (2002). Moreover, Dorgan and Dowdy (2004) put a numeric figure to the benefit of using IT *and* improved management practices: an increase of up to 20% in productivity is suggested to be attainable, but not if firms simply invest in IT without accompanying this investment by first-rate management practices. The study was conducted during 1994-2002 in 100 manufacturing firms located in the UK, the USA, France and Germany.

This recent research development contributes to increased understanding of how IT benefits productivity. Moreover, it provides a very welcome clarification amidst concerns – such as those voiced in the 1980s 'information technology productivity paradox' - that the expected IT impact on productivity would fail to materialise with due consistency. The productivity paradox may have been explained since then by the fact that IT investment mainly leads to higher product quality and variety, thus aggregate output does not reflect accurately the very costly and large-scale effort to improve technology. Instead, the existence of complementarities adds to the much-needed reasonable argumentation that IT has a positive impact on productivity when combined with the right mixture of management practices (Brynjolfsson *et al* 2000).

---

[1] Such as Brynjolfsson *et al* (2000), Black and Lynch (2001), Caroli and Van Reenen (2001), Bresnahan *et al* (2002), Dorgan and Dowdy (2004) or Battisti *et al* (2005)

On a final note, it is not yet clear whether the implementation of ICT would precede, trigger or follow (shortly) the implementation of complementary management practices in order for these synergetic complementary effects to be experienced. For instance, Caroli and Van Reenen (2001) show that the introduction of ICT seems to be associated with innovation and organisational change, leading to higher productivity. However, Battisti *et al* (2005), who also find positive complementarities between ICT and workplace innovation in their panel study of Italian plants, can not distinguish whether ICT precedes (and leads to) the adoption of workplace innovation or vice versa.

*JIT/ TQM and Productivity*

Just in time management (JIT) and Total Quality Management (TQM) are two management practices usually forming the pillars of coherent organisational systems initially inspired by Japanese production systems and aimed at maximising the speed of product delivery and service quality. JIT is an inventory strategy implemented to improve the ROI of a business by reducing in-process inventory and its associated costs. Although the foundations have been developed by Henry Ford in the early 1920s, the JIT philosophy became famous in the 1950s as part of the Toyota Production System. TQM is a set of customer-focused management strategies aimed at embedding awareness of quality in all organisational processes and thereby increasing customer satisfaction at continually lower real costs. Despite being at the origin implicitly aimed at increasing company efficiency, the results of the studies reviewed with regard to the impact of JIT and TQM on productivity are not conclusive. At firm-level, both JIT and TQM have been found to have mixed effects, ranging from positive, to none or even to negative effects (though the latter was only slightly significant and only in the case of JIT) (Callan *et al* 2000; Brox and Fader 1996, 1997 and 2002; Kaynak 2003; Kaynak and Pagan 2003; Sale and Imman 2003; or Callan *et al* 2005). Only one relevant plant-level study has been found in our review, and it reports a positive impact of JIT on productivity (Lawrence and Hottenstein 1995).

At firm-level, Brox and Fader (1996, 1997 and 2002) employ a generalised CES-TL cost model based on firm cost-functions in order to differentiate between the financial characteristics of JIT and non-JIT user firms, and find that JIT increases productivity and cost efficiency. JIT is defined as a mixture of JIT/TQM practices including Kanban, integrated product design, integrated supplier network, plan to reduce set-up time, quality circles, focused factory, preventive maintenance programs, line balancing, education about JIT, level schedules, stable cycle rates, market-paced final assembly, group technology, program to improve quality (product), program to improve quality (process), fast inventory transportation system, flexibility of worker's skills. This amalgamation of a large set of practices means that the impact of separate practices cannot be distinguished. Another slight drawback is that profitability or performance is measured as profit to investment, without being related to labour, unlike productivity, which is defined as labour productivity.

Lean production is another example of joint adoption of clusters of complementary principles. It originates from research into Japanese manufacturing (Womack *et al*, 1990). The basic principles are team-based work organisation, active problem solving, high-commitment HRM policies, lean factory practice, tightly integrated material flows, active information exchange, joint cost reduction, and shared destiny relations. Oliver *et al* (1996) analyse data collected from two international studies involving car component manufacturing companies in eight countries: France, Germany, Italy, Japan, Mexico, Spain, the UK and the USA. The questionnaire applied is designed specifically to facilitate the profiling of management practices to determine the extent of use of lean manufacturing practices. Using this single source of information, the study presents evidence that lean production principles partly explain high performance. Similarly, Lewis (2000), by means of a longitudinal study on lean production applied to the UK, France and Belgium, shows that lean production does not automatically result in improved financial performance. Indeed, being 'lean' can restrict the firm's ability to achieve long-term flexibility.

Similar results are found by Lawrence and Hottenstein (1995), Callan *et al* (2000), Kaynak and Pagan (2003) and Callan *et al* (2005), studies with the added benefit of allowing for a more refined management practice analysis. Lawrence and Hottenstein (1995) find a positive association between JIT and performance in their analysis of Mexican plants affiliated to USA companies. The study uses proxies for performance (quality, lead-time, productivity and customer's services) and for JIT management practices (extent of employees' participation, suppliers' participation and management commitment). Callen *et al* (2000) also find that JIT is associated with improved quality of process and product, lower costs and higher profits. It should however be noted that this study does not measure productivity per se, but profitability, defined as profit margin (operating profits divided by sales revenues) and contribution margin ratio (contribution margin divided by sales revenues). Kaynak and Pagan (2003) concentrate on estimating the JIT related sources of technical inefficiency, with results suggesting that internal organisational factors (such as the top management being committed to implementing JIT) are related to higher productivity, whereas external organisational factors (such as supplier value-added, or transportation issues) are not. Moreover, the study highlights that it is the degree of implementation of JIT, which is significantly related to performance, measured as financial and market performance, time-based quality performance, and inventory management performance). The study uses a stochastic frontier model for which the TL production function parameters are estimated simultaneously with the technical efficiency effects.

Subsequently, Callen *et al* (2005)'s ample study scrutinizes the interaction among performance outcomes, investment in JIT management practices, and productivity measurement at the plant-level, suggesting that productivity measurement mediates the relationship between performance outcomes and the intensity of JIT management practices adoption. The productivity measures used are TFP, labour productivity, ROI, quality of output, inventory (as total number of productivity measures associated with inventory control), and performance outcomes are measured via efficiency and profitability. A stochastic frontier production function of labour, capital, fuel and JIT technological index is estimated in order to obtain a correlation analysis between efficiency scores and plant profitability (i.e. EBIT/value of production at retail prices). Additionally, OLS and 2SLS estimators are used to model efficiency and profitability as a function of the JIT concentration index and the total number of productivity measures. The findings show that the broader the range of productivity measures, the more efficient and profitable the plants. Additionally, plants employing industry-driven productivity measures – especially if they are more JIT-intensive - are found to have higher profitability than those employing idiosyncratic productivity measures. Notably, even though plant profitability and efficiency are highly correlated, JIT-intensive plants tend to be more profitable but less efficient than their less JIT-intensive counterparts, which shows that JIT-intensive plants are still able to generate relatively higher profits despite leading to rather higher resource wastage.

Unlike the studies reviewed above, Sale and Imman (2003) combine the analysis of JIT adoption with the adoption of Theory of Constraints (TOC) in firms surveyed over a period of three years. Firms are categorised in four groups according to whether they adopt (1) only TOC, (2) only JIT, (3) both practices or (4) neither (traditional manufacturing). Firm performance level as well as performance change is followed. Performance measures are assessed using thirteen criteria weighted via management-reported importance scores, including sales level, growth rate, market share, operating profits, profits to sales ratio, cash flow from operations, ROI, new product development, market development, R&D activities, cost reduction programs, personnel development and political public affairs. Results from variance analysis show that TOC-only adopters achieved the greatest performance and improvement in performance, whereas JIT-only adopters did not have superior performance or superior change in performance when compared with traditional manufacturing. Lastly, firms using both JIT and TOC experience a drop in performance though this is only significant when compared against the TOC-only adopters.

Reports on the impact of TQM on performance are mixed. In one study, TQM exerts little or no observable effect on increasing productivity over the short time it was in place (Kleiner *et al*, 2002). In fact, it reduces labour productivity and increases labour costs, although a positive effect starts to be observed during the subsequent year. It is reasonable to expect that a time lag of some duration is required for a change in management practices to exert an impact, however this study offers initial insights that management under pressure for results are perhaps unable to commit to the achievement of long-term results if the short-term costs are too great. Instead, Kaynak (2003) reports evidence about the impact of TQM on firm's performance. Indeed, by using a combined sample of manufacturing and service firms, it shows a positive relationship between the extent to which companies implement TQM and firm performance. Three TQM practices (specifically: process management, supplier quality management, and product or service design) exert a direct effect on operating performance, and other TQM practices indirectly affect operating performance via those three practices. Operating performance mediates a positive effect of TQM practices on financial performance.

*HRM Practices and Productivity*

A strand of literature argues that investment in HRM practices can raise and sustain a high level of firm performance. HRM practices can represent a significant source of competitive advantage, as they are the means by which firms locate, develop and retain rare, non-imitable and non-substitutable human capital (Barney, 1991; Barney, 2001).

The studies found in the literature have predominantly reported a positive effect of using HRM practices although it needs to be ensured that costs for introducing and maintaining these practices do not outweigh their benefits. Empirical evidence suggests that unionisation is an important mediator for the success of HRM practices.

Koch and McGrath (1996) investigated the impact of a set of HRM practices on labour productivity, to find that investments in HR planning and in hiring practices are positively associated with labour productivity. Results suggest that firms that systematically train and develop their workers are more likely to enjoy the rewards of a more productive workforce than those that do not, although this is not framed to take account of the bigger picture. For example, Capelli and Neumark (2001) provided some indication that empowering work practices are related to greater productivity. The authors presented partial evidence of such a relationship, however, since the work practices raise labour costs per employee (in this case employee compensation), it is unclear whether such practices are beneficial to the firm overall. Another study, this time of small Belgian companies, revealed a similar situation. Sels *et al* (2006) demonstrated a strong and positive relationship between HRM intensity and productivity, controlling for past performance and using one-year lagged financial performance indicators (although the measures were recorded contemporaneously). This beneficial effect was greatly outweighed by the cost increases associated with higher HRM intensity. Nevertheless, HRM intensity was directly related to profitability, and the authors understand this in terms of the minimisation of unmeasured operational issues.

A cross-sectional, single-respondent empirical study of 52 Japanese multinational corporation subsidiaries in the US and Russia demonstrated that employee skills, attitudes and behaviours play a mediating role between HR systems and firm outcomes (Park *et al*, 2003). Results suggest that clusters of HR practices positively influence the performance of the types of Japanese subsidiaries concerned. This can be explained in one of two ways: either HRM practices exert an influence regardless of firm location, or Japanese organisations always implement very similar 'best practices'. Indeed, other empirical evidence suggests that the potential causational path from HRM practices to productivity is more complicated than once thought. Another study, multi-respondent and quasi-longitudinal in design involving Indian software companies presented empirical evidence demonstrating no direct causal relationship between the HRM practices in question and organisational financial performance, although some HRM practices were directly related to operational performance parameters (Paul and Anantharaman, 2003). Instead, it was found that every single HRM practice measured indirectly influenced the organisation's operational and financial performance. The indirect effect is very important, because few studies employ a research design where intervening variables are measured, but beware that the sample size was too small to apply all of the desired statistical analyses (i.e. maximum likelihood model) and no controls were added. The findings are nevertheless thought provoking and infer that simply focussing on a direct linkage between HRM and performance may not reveal the operational mechanism through which an effect is exerted.

In support of this type of approach, Michie and Sheehan (2005) analysed original data from a mixed sample of 362 manufacturing and service sector companies. The empirical findings demonstrate positive relationships between HR policies and practices and objective financial performance, mediated by business strategy type (business strategies were classified as cost leadership, innovation-focused or quality-focused). Additionally, the use of external flexible labour was associated with lower HR effectiveness. The implications are very pragmatic, and although this survey is only cross-sectional, it could be inferred that there exists a two-way causational relationship between the HR policies and practices and financial performance.

Ichniowski *et al* (1995) formed a statistical distribution of HRM practices to show that some practices are adopted only in presence of some others (i.e. as clusters), and some clusters display a more significant productivity advantage than others. The econometric analysis of this paper is relatively robust as it is based on panel data rather than on cross-sectional data. Building on the findings, Ichniowski *et al* (1997) analysed the impact of different clusters of management practices on productivity, to estimate the impact of a single HRM practice on productivity. Empirical results demonstrated that manufacturing lines using a set of HRM practices are associated with a higher level of productivity than lines employing a single HRM practice.

High Involvement Work (HIW) practices represent another important set of HRM practices. Employees of a high-involvement organisation take greater responsibility for its success. In practice, this involves HRM practices to develop and support a self-managing and self-programming workforce (Lawler, 1992). Guthrie (2001) received responses from 190 New Zealand companies with at least 100 employees, and empirically demonstrated a positive relationship between the application of high-involvement work practices and productivity. However, an interaction was observed with employee turnover: when productivity was high, turnover was linked to decreased productivity; and when productivity was low, turnover was associated with increased productivity. Indeed, employee retention is critical when financial investments in work practices are relatively high, and this finding infers that employers may benefit from utilising complementary management practices (such as enhancing retention of good performers) alongside high-involvement systems.

Bryson *et al* (2005) investigated WERS98 data (a nationally representative sample of organisational data collected using a preferable technique of multi-respondent sampling across organisations) for the private sector only to test hypotheses regarding work organisation, trade union representation and workplace performance. Findings demonstrated a positive effect of HIW practices on labour productivity; however, this effect was minimal within non-unionised workplaces. Descriptive evidence suggests this effect is attributable to concessionary wage bargaining on the part of unions.

When comparing the productivity of Japanese and USA production line workers, empirical evidence shows that USA manufacturers who had adopted a full system of innovative HRM practices patterned after the successful Japanese system, achieved levels of productivity and quality equal to the performance of Japanese manufacturers (Ichniowski and Shaw, 1999). This suggests that the higher average productivity of Japanese plants cannot be attributed to cultural differences; instead, this is related to the utilisation of more effective HRM practices.

*Joint Adoption of Operational and HRM Practices*

It seems that there is consensus in the literature about a positive impact of an individual management practice in isolation on productivity. It is also worth mentioning that across the extant literature this is the most common approach of investigation.

However, recent theoretical and empirical research suggests that this approach may be misleading since firms often adopt clusters of management practices rather than individual practices in isolation (Ichniowski *et al*, 1995; Huselid, 1995; Patterson *et al*, 2004; among others). This is because the presence of complementarities among innovations is such that when an innovation is adopted in isolation it might not necessarily yield positive gains. However, when innovations are jointly adopted they can significantly improve productivity, increase quality and often result in better firm performances than more traditional systems (see for example Ichniowski *et al* (1997) and Ruigrok *et al* (1999) for applications to HRM practices or Stoneman (2004) and Battisti *et al* (2005) for theoretical models). In other words, the benefits from the joint adoption of clusters of complementary innovations can be higher than the sum of the individual effects.

Other studies at firm-level are sceptical about the positive effect of joint adoption of management practices. Patterson *et al* (2004), for example, analyse the impact of a cluster of management practices upon performance by taking into account the possible complementarities between OM and HRM practices. Thus, by distinguishing between integrated manufacturing (i.e. OM) and empowerment (i.e. HRM) practices, this study uses multiple regression analysis to test the following three key assumptions: whether OM practices affect HRM practices, whether OM practices and HRM practices enhance the company performance and whether there is interaction between OM and HRM practices. The empirical results seem to challenge the common view that management practices may affect firm productivity/performance. They show that there is no relationship between integrated manufacturing and empowerment practices and the study did not find any evidence in support of a relationship between the impacts of OM practices upon firm performance. This result questions the findings of the most part of literature and casts doubts on the ability of management practices to affect positively the firm performance.

Birdi *et al* (2006) investigated the relationships over time between the introduction of seven OM and HRM practices (JIT, TQM, AMT, supply-chain management, empowerment, learning culture and teamwork) and audited company performance for 308 companies over a period of 22 years. Results demonstrate a universally positive effect of empowerment on performance, whereas the impact of learning culture appeared to be context-specific. Importantly, the impact of the other five practices varied, indicating that the introduction of a particular management practice can have no or even a negative impact on performance. Statistical relationships between variables were largely incompatible with contemporary theories; a significant finding given that it is highly unlikely these propositions have been previously tested on such a grand scale. Given the single respondent design, the authors conducted a consistency check and yielded a high consistency rate (84%). It is difficult to criticise this study due to the exceptionally extensive data set and explicit methodology, although the extent of implementation of each practice is not ascertained and it is unclear whether the cessation of practices is incorporated in the analyses. The authors argue that it is likely that only effective practices are institutionalised by an organisation and consequently reported as in use; however, this relies upon the assumption effective feedback mechanisms exist to provide accurate information to the organisation's decision-makers.

Bloom and Van Reenen (2006) collected data on 732 manufacturing firms in the UK, France, Germany and the USA for the period 1992-2004. Data collection involved the application of a novel measurement tool, offering a sophisticated way of assessing and combining ratings of OM and HRM practices at a grassroots-level. Robust estimation techniques were applied (specifically OLS, IV, WG and GMM estimators). The resulting measures of managerial best practice are strongly associated with sales growth, survival, Tobin's Q, profitability and productivity. The authors investigated why so many companies survive with relatively inferior management practices, and why this pattern varies so much across the USA and Europe. Findings suggest these phenomena can be explained in terms of low product market competition and eldest sons inheriting control of the family firm. Both of these factors are much more prevalent in the European countries surveyed than the USA, and accounted for around half of the badly managed firms and a similar amount of the inter-continental discrepancy in management performance. The authors also uncovered a large variation of management practices even across firms within each country, especially for the UK. The methodology of this study is commendable and many different variables are controlled for. However, the universal conceptualisation of particular practices as 'good' or 'bad' provides only a proxy of management practices and does not allow for the incorporation of context-specific practices that may be more important to other sectors aside from manufacturing.

*Discussion and Conclusions*

In the survey, we have focused the attention on the relationship between OM and HRM practices and productivity. We have observed, on the one hand, how these management practices have been measured and, on the other hand, how the impact of these practices on productivity has been estimated. From the literature, we have found that there is consensus amongst researchers on a generally positive effect of individual management practices on productivity or performance when considered in isolation. However, when management practices are jointly adopted, there is no consensus on a positive effect. Furthermore, we have noticed that although the econometric methodology appears to be robust and quite sophisticated, a wide range of definitions of management practices, productivity and performance have been used, which makes results not robust to comparisons over time and across studies.

These results have some important implications. Our findings suggest that the lack of consensus over (a) the definition, (b) the measurement and (c) the level of analysis of management practices is a principal reason behind the wealth of contradictory reports on complementary management practices. Indeed, the adoption and implementation of complementary practices is found to have effects that vary in sign and size, depending on the definition and measurement of the studied management practice and of performance. Additionally, data collected is often based on a simplistic and subjective analysis of the extent to which management practices have been adopted and implemented, which then hinders researchers' attempts to generalise findings. For instance, the questionnaire may only ask whether training has occurred in the organisation, prompting a yes/no answer, whereas further in-depth measures of the amount and type of training would be more informative and lead to less-biased research findings. Moreover, variations in the level of analysis account for further research difficulties in making comparisons. Data typically available comes from cross-section studies performed at industry – predominantly in manufacturing - or firm level studies. However, in order to understand organisational changes that may take some time to become apparent, longitudinal data as well as plant or establishment level data would be much more appropriate.

The prevalence of correlational studies indicates that many researchers are at an early, exploratory stage of trying to understand the mechanics behind how management practices may influence productivity. This type of research design does not facilitate the inference of causality, and is extremely limited in the way it can convey the complexities of relationships between people and processes. Cross-sectional research designs test simultaneous effects, i.e. two-way causal relationship between two variables. A fair number of studies are also limited by small sample size, reducing external validity. Indeed, there are serious concerns about the methodological limitations of research into a link between management practices and productivity (for a thorough review see Wall and Wood, 2005).

Some studies have adopted longitudinal designs with varying success. Indeed, it is more reasonable to conclude that there needs to be some kind of time lag between initial implementation, employee consultation, or union negotiation and the management practices demonstrating some kind of impact on organisational outcomes. It is important to mention here the potential reverse causality of management practices (Savery and Luks, 2004).

The majority of research reviewed has relied upon data collected from single respondents, increasing the chances of common method variance. Undoubtedly, there is an inherent trade-off between reducing common method variance associated with single-respondent designs and ensuring a large enough sample size and sufficiently high response rate to draw generalisable conclusions. It is important to balance the needs of good science with more pragmatic concerns, and appropriate statistical tests can be applied to test for bias prior to subsequent analyses, for example see (Birdi *et al*, 2006).

Many studies have also relied entirely upon perceptual measures that may incorporate measurement error. However, Wall *et al* (2004) empirically demonstrated that perceptual measures of company performance are no less valid or reliable than objective measures. Indeed, there is an argument against using company accounts: accounting conventions and other sources of error may pervade this assumed objective data. It is possible that purely financial performance measures fail to account for the broader organisational picture, therefore the inclusion of non-financial performance criteria such as customer satisfaction, productivity and quality provide may provide outcomes that are more amenable. To the contrary, if the bottom line contribution of management practices cannot be demonstrated then their implementation remains highly questionable. A small number of key studies have demonstrated promising linkages between management practices and financial performance (Michie and Sheehan, 2005; Paul and Anantharaman, 2003).

In general, the multi-dimensional nature of management practices translates into a complex relationship between them and productivity measures. Empirical evidence suggests that effective management practices need to be context specific, as productivity indices need to reflect a particular organisation's activities. Consequently, it is tricky to ascertain whether the finding of a relationship, or no relationship, is a fair conclusion. Some researchers have risen to the challenge and adopted more sophisticated methods of operationalisation and analysis. For example, Bloom and Van Reenen (2006) offer greater scope for unravelling the complex interrelated and mediationary relationships at play. Another study uncovered a curvilinear relationship between management practices and performance (Maes *et al*, 2005), indicating that beyond a certain amount or intensity management practices actually diminish performance. Correlational research to ascertain relationships between other workplace constructs and productivity may help inform future research into mediation, such as Geralis and Terziovski (2003) or Silvestro (2002).

There is a fair amount of support for a contingency approach; however, it is unclear what the common factors to consider are, see Birdi *et al* (2006). Applying context-specific measures creates variability between research findings and renders them directly incomparable. For example, it is apparent there are contrasting definitions of lean production techniques, and these difficulties in achieving consensus makes it likely that each firm follows a 'unique lean production trajectory' (Lewis, 2000; p. 975). Whereas on the other hand, TQM practices tend to be involve a similar set of practices within whichever organisation they are implemented within. Indeed, there remains scope for the future investigation of degrees of internal, organisational and strategic fit (Wall and Wood, 2005).

For this and other reasons, we strongly believe that there is need for further research. In particular, for a multi-level approach from the lowest possible level of aggregation up to the firm-level of analysis in order to assess the impact of management practices upon the productivity of firms. When the research is conducted it should always be considered that, what is most important is not the introduction of the management practice but rather how it has been introduced, when it is introduced and how it has been implemented (this issue has been examined by Ichniowski *et al* (2003) and Leseure *et al* (2004), amongst others).

**Appendix - Six examples of management practices measurement schemes**

1. A questionnaire is sent to a pool of firms, to ask them (with 1= not used at all; and 10= used to a large extent):
   - The degree of use of practices
   - The importance attached to some performance criteria
   - The degree of satisfaction of the top management about the performance of each criteria.

2. Management practices or innovations are defined not as technological innovations but as improvements in the way things are done. In this context as: a) degree of decentralization leading to a high level of workers responsibility, including for example, responsibility for quality control and team based production, b) strong incentives for individual performance, large profit related bonuses, promotion and job security; c) small number of job classification; d) extensive screening of prospective employees; e) close and continuing relationships with suppliers and JIT scheduling practices.

3. Management practices or innovations are measured by asking managers whether there have been in the firm:
   - Reductions in restrictive practices by employees
   - Introduction of new technologies
   - Changes in the organizational structure
   - Increases in decentralization
   - Adoption of new human resources management practices
   - Changes in the industrial relations
   - The initiation of new JIT practices

4. A survey instrument is designated to collect the data about JIT. Firms were contacted by telephone, then visited and interviews took place with plant managers or production managers or an owner/CEO: The firms are classified as JIT users not only according to a self-declaration of being JIT users but also according to 17 management strategies (Kanban, Integrated product design, Integrated supplier network, Plan to reduce setup time, Quality circles, Focused factory, Preventive maintenance programs, Line balancing, Education about JIT, Level schedules, Stable cycle rates, Market-paced final assembly, Group technology, Program to improve quality (product), Program to improve quality (process), Fast inventory transportation system, Flexibility of worker's skills) designated by the survey to capture the extent of JIT use.

5. A five point scale is used to measure how well companies have implemented three important management practices: Lean manufacturing (which cuts wastes in the production process), Performance management (which sets clear goals and rewards employees who reach them) and Talent management (which attracts and develops high-caliber people).

6. Direct measures of management practices: For example, in the case of ICT is used a precise variable to capture the amount of IT investment. Sometimes, similarly happens in the case of HR management practices: for example skills are measured for firms with the exact proportion of variables designated to capture the employees' skills. Employee Involvement for example is measured as the percentage of all employees significantly impacted by Employee Involvement programs. TQM as the percentage of employees impacted by a TQM program, etc.